# What sustains stable cultural diversity and what undermines it? Axelrod and beyond[1]

**Andreas Flache\*, Michael W. Macy[+]**

\*Department of Sociology, University of Groningen, a.flache@rug.nl
[+]Department of Sociology, Cornell University, mwm14@cornell.edu

**Abstract**
Despite fifty years of empirical and theoretical research on social influence, one of the great puzzles of social science remains unanswered: How is cultural diversity possible? Axelrod (1997) has proposed a model of cultural dissemination that explains cultural diversity as the consequence of two fundamental social mechanisms – social influence and homophily. Subsequent work has shown that within this framework, cultural diversity is not robust against small amounts of noise. However, a key simplification of the original analysis has not been explored – Axelrod's assumption that all cultural states are nominal. We integrate metric states into Axelrod's original model and thus align it with previous formal models of social influence that did not incorporate homophily. We show that in the extended cultural dissemination model, metric states undermine cultural diversity, even without noise, by creating sufficient cultural overlap for mutual influence between agents. We then show how "bounded confidence," a recent extension of social influence models, allows cultural diversity to persist in cultural dissemination models as well, even in the presence of noise. However, the solution turns out to be less robust than we would like. Diversity can be sustained only with a relatively small number of metric states, low levels of noise, and a narrow confidence interval. We then suggest the need to incorporate "negativity" into future studies as a possible solution to the puzzle of cultural diversity.

## 1. Introduction

Cultural diversity is both persistent and precarious. People in different regions of the world are increasingly exposed to global influences from mass media, internet communication, interregional migration and mass tourism. While some authors expect that this may lead to the emergence of a global monoculture, others point to the pervasiveness of cultural differences (cf. Greig, 2002). How is stable diversity of cultures possible? Recently, formal modelers of cultural dynamics, most prominently Axelrod (1997), have argued that a few simple, fundamental mechanisms of social interaction may suffice to explain stable cultural diversity and its conditions (e.g. Latané and Nowak, 1997; Mark 1998; Klemm et al 2003a,b; Macy et al, 2003). These models combine two elementary social mechanisms, homophily and social influence. *Homophily* is attraction towards similar others, or the law that "birds of a feather flock together" (McPherson et al, 2001). *Social influence* implies the tendency to become more similar to influential others (Festinger et al, 1950). It seems straightforward that a combination of social influence and homophily drives populations towards cultural homogeneity. However, Axelrod's (1997) computational studies showed how local convergence can lead to global differentiation. The key assumption that generated this result was cultural speciation. Axelrod assumed social interaction between neighbors to be entirely cut off when actors disagree in all cultural dimensions, analogous to the inability of sexual organisms with a common ancestor to mate once they differentiate beyond a critical threshold. As a consequence, homophily and social influence can generate isolated regions that are stable, because there is no more influence from neighbors and agents share the same culture within the region.

Axelrod's result has inspired a range of follow up studies, many of which supported his basic conclusions (e.g. Mark 1998, 2003; Parisi et al 2003), albeit with some modifications (e.g. Shibani et al 2001; Greig, 2002). But Klemm et al (2003a,b; (for a recent overview see San Miguel et al, 2005) challenged the robustness of stable diversity in Axelrod's model. They showed that a population that exhibits stable diversity under Axelrod's assumptions "drifts" into a virtual monoculture if a very small amount of "cultural mutation" (or noise) distorts the dynamics. This resonates with earlier results obtained using models of opinion dynamics in influence networks with fixed structures and without the assumption of homophilous dynamic networks (French 1956; Abelson 1964; Friedkin & Johnson 1999). Abelson proved that for these fixed-network opinion formation models, a very weak condition is sufficient to guarantee convergence to global consensus, namely, the network needs to be "compact," such that there are no subgroups that are entirely cut off from outside influences.

[1] This research has been supported by the Netherlands Organisation for Scientific Research, NWO, and by the U.S. National Science Foundation program in Human Social Dynamics (SES-0432917). Our work has benefited from stimulating discussions with Károly Takács, Michael Mäs and other members of the discussion group on norms and networks at the Department of Sociology of the University of Groningen.

In this paper, we will show that conditions for stable diversity become even more restrictive when we relax another implausible assumption in Axelrod's original model, the assumption that all cultural differences are qualitative (hence represented as nominal scales). This is a key difference between opinion formation models (in the tradition of French and Abelson) and the cultural dissemination framework. In the wake of French and Abelson, opinion formation models employ metric state spaces, which allow for the gradual and stepwise convergence towards uniformity that is the typical model outcome. In this paper, we will integrate into Axelrod's cultural dissemination framework the assumption that at least some cultural dimensions are metric. We show that – as in traditional social influence models – cultural homogeneity is then the ineluctable outcome. Moreover, that is the case even when we assume away all noise. Given that different groups almost certainly have at least one cultural difference that is quantitative (such as how much they like jazz), we are back to our original question: how is stable diversity possible?

We propose and analyze a solution to the puzzle. We adopt from Deffuant et al. (2000, 2005) a recent extension of opinion formation models. Their assumption of "bounded confidence" means that agents are only influenced by those whose opinions differ from their own by not more than a certain threshold (see also Hegselmann & Krause 2002). Bounded confidence implements Axelrod's original homophily assumption, but now for a metric state space. Recent studies using fixed-structure social influence models have shown how bounded confidence can explain stable diversity. With moderate confidence thresholds, differentiation into a small number of opposed radical factions becomes possible. We conjecture that bounded confidence also leads to stable diversity in the cultural dissemination framework (with homophilous dynamic networks). To test this, we incorporate bounded confidence assumptions into our extended version of the cultural dissemination model. We then systematically explore the effects of different confidence thresholds on stable cultural diversity in state spaces that include both metric and nominal states. Section 2 describes our model. In Section 3 we present computational experiments and results. In section 4, results are discussed and conclusions are drawn.

## 2. Model

The simulated population consists of $N$ agents, of which each agent has a subset of the population as his neighborhood. Axelrod used a regular lattice to spatially order the population. We are not interested in effects of the spatial arrangement in this paper, so we adopt Axelrod's regular lattice with edges (no torus) and von Neumann neighborhood of size 4. Axelrod typically used grids with 10x10 cells, but also studied different grid sizes. Every cell is occupied by exactly one agent. At any point in time, the cultural state of an agent $i$ is a vector of $F$ elements, called "features". On any single feature, an agent has a "trait" represented by an integer value in the range of $0..Q-1$, where $Q$ is the number of possible traits. Formally, the state of an agent is

$$s_i = (s_{i1}, s_{i2},...,s_{iF}), \quad s_{ix} \in \{0,1,...,Q-1\} \subset N_0. \quad (1)$$

In the original cultural dissemination model, the distance between two traits on the same feature is meaningless. The traits of two different agents on this feature are either equal or they are different. To integrate metric features, we assume that for metric features distances between two traits $q$ and $r$ in this range are meaningfully defined by $q-r$. To distinguish metric from nominal features, the parameter $F_n \leq F$ indicates the number of nominal features. All features $f$ for which $f \leq F_n$ are nominal (as in the original model), all other features are metric. This model is equivalent to the original cultural dissemination model if $F_n=F$.

The cultural dissemination model specifies dynamics as follows. Initially, every agent is assigned a random state where every possible trait value has equal probability to be assigned. In every discreet time step $t$, one pair of neighbors ($ij$) randomly gets an interaction chance with equal probability. Following Klemm et al (2003a,b), we assume a small probability $r$ that in one time step prior to interaction, the trait value of a randomly selected agent on a randomly selected feature will be perturbed to a randomly selected new value. When selected, the probability for actual interaction between $i$ and $j$, $p_{ij}$, is derived from the cultural overlap $o_{ij}$ between $i$ and $j$, where $0 \leq o_{ij} \leq 1$. In Axelrod's original approach, cultural overlap is computed as the proportion of features on which $i$ and $j$ have identical traits. To adapt this for metric features, we define cultural overlap between two agents $i$ and $j$ for a metric feature $f$, $o_{ijf}$, as distance between their traits on this feature, divided by the maximal possible distance ($0 \leq o_{ij} \leq 1$). To obtain a comparable overlap measure for nominal features, we set $o_{ijf}=1$ if $i$ and $j$ have different traits on this feature, and $o_{ijf}$ 0 otherwise. Technically,

$$o_{ijf} = \begin{cases} 1 - \dfrac{|s_{if} - s_{jf}|}{Q-1}, & \text{if } f > F_n \\ 0, & \text{if } f \leq F_n \text{ and } s_{if} \neq s_{jf} \\ 1, & \text{if } f \leq F_n \text{ and } s_{if} = s_{jf} \end{cases} \qquad (2)$$

The overall cultural overlap between $i$ and $j$, $o_{ij}$, is obtained as the average cultural overlap across all features, or $o_{ij} = 1/F \sum_{f=1}^{F} o_{ijf}$ (3). This definition of overlap is compatible with Axelrod's original model when $F = F_n$.

To implement bounded confidence, we assume that the probability of interaction, $p_{ij}$, is a non-linear function of cultural overlap, unless confidence is not bounded. The probability of interaction increases linearly in cultural similarity, but only when cultural overlap exceeds the confidence threshold $\tau$, where $0 \leq \tau \leq 1$. We also assume that the probability of interaction may be affected by a small amount of random interaction. Klemm et al have shown how mutation can considerably change the dynamics of the model, because it can create bridges for influence between agents who are otherwise too dissimilar to interact. However, random interaction may also occur without the spontaneous change of traits. In many social contexts, such as culturally mixed classrooms or work groups, at least occasional social influence takes place between culturally dissimilar people, without the need of a change of their traits prior to interaction. To capture this random interaction, we introduce a second noise parameter $r'$ that enters into the calculation of the actual probability of interaction. When the cultural overlap falls below the confidence threshold, the probability of interaction is $r'$. Technically,

$$p_{ij} = c(o_{ij}), \text{ where } c(o) = \begin{cases} r', & o \leq \tau \\ o, & \text{otherwise} \end{cases} \qquad (4)$$

Equation (2), (3) and (4) implement the assumption of homophily in the extended model of cultural dissemination. If the confidence threshold is set to its maximal level, i.e. $\tau = 1$, and both noise parameters are zero $p_{ij}$ is defined equivalently to the original model. Social influence is likewise implemented equivalently to Axelrod (1997). When interaction takes place and there is at least one feature on which $i$ and $j$ do not yet agree, one of them is selected at random to adopt the position of his counterpart on a randomly selected feature in which they are different. The extended model reduces to Axelrod's specification, if $\tau = 1$, $F_n = F$, and $r = r' = 0$.

## 3. Results

### 3.1 Replications

We begin by replicating Axelrod's (1997) model in which there is no noise ($r = r' = 0$), a purely nominal state space ($F = F_n$), and unbounded confidence ($\tau = 1$). Results show that under a large range of conditions the following pattern arises: Initially, cultural diversity declines rapidly and the average similarity between neighbors increases. At some time $t$, the system reaches an equilibrium state with stable diversity, i.e. with a number of distinct "cultural regions" in the population. A cultural region is a set of direct or indirect neighbors who all have the same cultural state. Axelrod (and many follow up papers) used the number of cultural regions in equilibrium as a measure of diversity[2]. To provide a baseline and to validate our re-implementation of the original model, we used the parameter settings ($\tau = 1$, $F_n = F$, and $r = r' = 0$) to replicate a condition that Axelrod showed to be conducive to relatively high diversity. In this *baseline condition*, $N = 100$ (10x10 grid), neighborhoods are small (4 neighbors), there are few features ($F = 5$) and there are relatively many trait values per feature ($Q = 15$). Based on 10 realizations, Axelrod reported for this condition an average of 20 stable "cultural regions" in equilibrium. We obtained an average of 19.55 based on 100 realizations[3].

---

2 Klemm et al (2003a,b) use the relative size of the largest cultural region in equilibrium.

3 We also replicated all other results reported in Axelrod (1997) very closely. The analyses reported in this paper start from a baseline condition that Axelrod has shown to be conducive to high diversity under the assumptions of his original model. We have also explored other sets of conditions (in particular more features and traits, other spatial and neighborhood structures). Results of these studies generally support our conclusions and will be reported in a longer version of the paper.

We then replicated Klemm et al. (2003a) by introducing a small amount of mutation ($r >0$). As expected, we found that small levels of mutation ($r \leq \approx 10^{-3}$) were sufficient to destabilize diversity so that the system gradually moves from a random start towards a metastable state close to cultural homogeneity.

Figure 1 shows a typical dynamic generated by our implementation for a noise level of $r = 10^{-3}$ ($r' =0$). Figure 1 demonstrates how even with mutation, cultural differentiation can persist for a long time. Such states typically persist for 100000 or more time steps in this condition, and then change in a relatively rapid transition into a new, temporarily stable state. Typically, cultural diversity declines in the course of these transitions and the population moves towards a near monoculture. For reliability, we ran 20 independent realizations of this experiment until the level of diversity, averaged over 200.000 iterations, became stable or the number of iterations exceeded $10^7$. We found an average of 1.55 cultural regions in the end-state. Figure 1 also illustrates how, once a near-monocultural state is reached, there can be some fluctuation that quickly moves the population out of this state into a new consensus state (between $t$=340000 and $t$=390000 in the example). This shows that mutations have two effects. First, new states can arise spontaneously and spread. Second, mutations can create overlap and thus bridges for social influence between otherwise isolated cultural regions, reducing diversity in the system. At the low level of noise used for figure 1, the second effect prevails. Mutations prevent the population from "freezing" in multicultural states with multiple disconnected cultural regions, with the eventual result of a monoculture with small random distortions. The mechanism of random interaction is different from mutation. Random interaction only creates bridges for influence between otherwise disconnected actors, but does not also introduce new diversity in the population. The long run outcome of random interaction can only be monoculture. To explain, monocultures are the unique absorbing states of the dynamics and from every multicultural state, several monocultural states are reachable within a finite number of steps. Additional experiments confirm this expectation, but show also that $10^7$ iterations are not sufficient to reach monoculture in all realizations. A ceteris paribus replication of the baseline condition with random interaction ($r' =0.001$) but without mutation ($r=0$) yielded on average 1.11 stable cultural regions in the end-state.

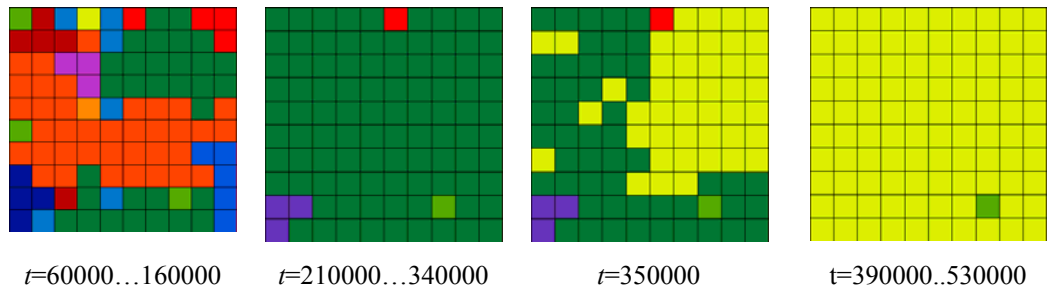

$t$=60000…160000    $t$=210000…340000    $t$=350000    t=390000..530000

*Figure 1*. Dynamics of typical model run with mutation ($N$=100, von Neumann neighborhood, $F$=5, $Q$=15, $\tau$ = 1, $F_n$=$F$, and mutation, $r$=0.001, $r'$ =0). Color of a cell represents trait on feature 1.

**3.2 Metric features**
Axelrod (1997) and most follow up papers use a concept of polarization that does not distinguish between dissimilarity and antagonism. In their terminology, a "polarized" outcome is a multicultural state in which neighboring cultures are different on all features. Cultural features are assumed to differ qualitatively and are modeled with nominal scales. Two agents either agree or disagree on a certain feature, but there is no *degree* of agreement or disagreement on this feature. This differs from many empirical studies (e.g. DiMaggio, Evans, and Bryson 1996), who define polarization as the tendency for individuals to hold extreme and uncompromising views. More importantly, maximal disagreement, the only condition under which two neighbors no longer influence each other in the cultural dissemination model, is a much more unusual and extreme situation when states are metric. With metric features, maximal disagreement requires that two neighbors adopt traits at opposite ends of the scale on every feature. Moreover, with a nominal feature, a change from total agreement on a feature can only be towards total disagreement, resulting in an overlap of zero after the change. By contrast, with a metric feature change can decrease the level of agreement, but leaves overlap above zero except for extreme cases. Accordingly, we expect that – ceteris paribus – there will be considerably less diversity compared to Axelrod's original model with all nominal features if at least one cultural feature is metric. As a test, we replicated the baseline condition, but with only a single change, the introduction of one metric feature ($F_n$ =4). Figure 2 below shows a typical dynamic for this scenario.

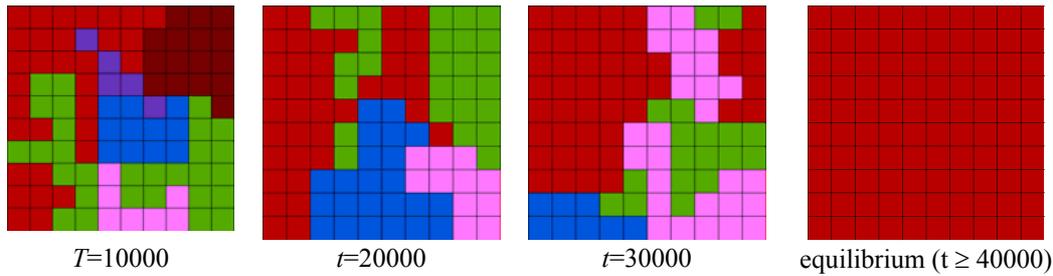

| $T$=10000 | $t$=20000 | $t$=30000 | equilibrium (t ≥ 40000) |

*Figure 2*. Dynamics of typical model run with one metric feature ($N$=100, von Neumann neighborhood, $F=F_n$ =5, $F_N$ = 4, $Q$=15, $\tau$ = 1, $r=r'$ =0). Color of a cell represents trait on *nominal* feature 1.

Figure 2 reveals a striking effect of metric features. Only one metric feature is sufficient to entirely wipe out diversity in a condition that is highly conducive to stable diversity without metric features. No noise is needed for this ($r=r'$ = 0). The explanation is that with a single metric feature, it is much less likely that cultural overlap between any two agents declines to zero. As a consequence, direct or indirect influence remains possible between all agents in the population. Sooner or later, all agents will thus come to agree with all their neighbors on all features, including those that are nominal. This reflects the results of the social influence models with static networks, but it also shows that unlike in these models, the assumption of a *continuous* opinion space is not even needed to obtain monoculture. With $Q$=15, our opinion space is discreet. It is the metric that matters, not the continuity. For reliability, we ran 100 independent realizations. In all realizations, the population converged to full homogeneity in equilibrium.

We found that one metric feature is sufficient to eradicate diversity in the baseline condition, but what about other conditions? As a more extreme test, we further reduced the relative weight of the metric feature by increasing the number of nominal features to 49, with just one metric feature. However, as Axelrod showed, more features also decrease diversity in the absence of metric features. To control for this, we simultaneously set $Q$=150, ten times as much as in the baseline. As a benchmark, we ran the experiment without a metric feature and with 50 nominal features and 150 traits ($F=F_n$= 50, $Q$=150, $r=r'$ =0, $\tau$=0). Based on 20 realizations, we obtained 15.8 average stable cultural regions in equilibrium. However, with one metric feature ($F$=50, $F_n$ = 49, ceteris paribus), we found that monoculture is the unique equilibrium. This shows that a single metric feature undermines diversity even when the proportion of features that are metric is negligible.

**3.3 Metric features and bounded confidence**
Metric features eliminate stable diversity for a simple reason: neighbors are rarely maximally dissimilar and therefore retain a positive probability of interaction, however small it may be. Using a solution similar to the one proposed by Axelrod, recent studies using fixed-network "opinion formation" models showed how stable diversity can nevertheless obtain even with metric opinions. "Bounded confidence" models assume that agents are influenced only by those other agents whose opinion differs from their own not more than a certain confidence level (e.g. Hegselmann and Krause, 2002; Deffuant et al, 2005). These studies show that moderate confidence levels are sufficient for polarization of the population into a small number of factions.

However, these studies used the "opinion formation" model where networks are assumed to be fixed and only the states of nodes (opinions) can vary. It is not at all obvious that this result generalizes to dynamic networks in which relations vary in response to changes in the nodes. We want to know if bounded confidence can also explain diversity in "cultural dissemination" models with network dynamics driven by homophily.

There are two potential problems in generalizing from the bounded confidence results in previous studies -- the restricted size of the confidence intervals (i.e. low confidence thresholds) and the absence of noise. Bounded confidence studies typically assume much more restricted confidence levels than those that correspond to Axelrod's model with nominal features. For example, Hegselmann and Krause (2002) assumed 5% confidence intervals around each opinion, within which neighbors could interact. That is an extremely narrow interval compared to what Axelrod assumed. Using nominal features, Axelrod was able to obtain stable cultural regions with $F= F_n$ =5, which corresponds to a cultural difference of 80% of the maximum (i.e., neighbors have different traits on 4 out of 5 features). We need to know if metric features with 80% confidence intervals can also sustain cultural diversity, even without noise. In addition, allowing for noise (either as mutation or random interaction) seems likely to destabilize diversity, even with highly restricted confidence intervals.

To answer these questions, we conducted three computational experiments in which we manipulated the size of confidence levels and the amount of noise. In the first experiment, we assessed the maximum interval on a metric feature that is possible before diversity collapses, without allowing for any noise. We then ran two ceteris paribus replications, one with only mutation ($r=0.001$, $r'=0$) and one with only random interaction ($r=0$, $r'=0.001$). In all experiments, we started from a model with a very narrow interval ($\tau=0.05$) and increased the confidence interval up to unbounded confidence ($\tau=1$), in steps of 0.05. All other assumptions remained unchanged relative to the baseline condition with one metric feature ($F=5$, $F_n=4$, $Q=15$). Figure 3 charts the effects of $\tau$ on the average number of cultural regions in equilibrium, based on 20 realizations in each condition.

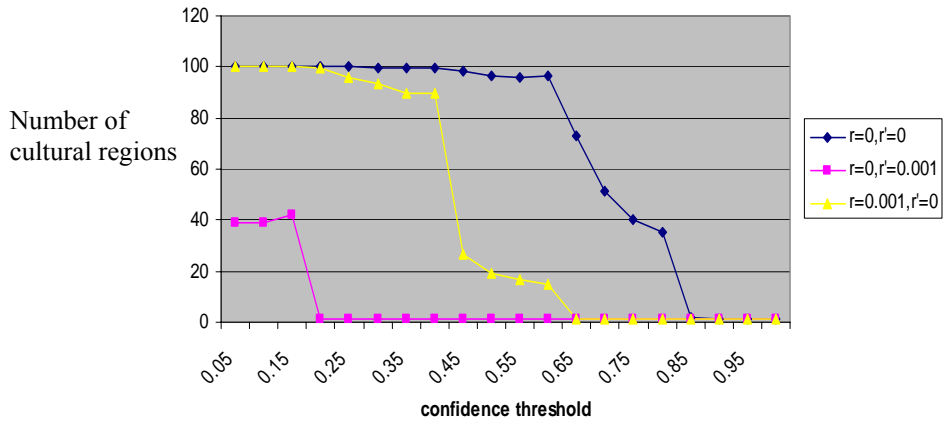

*Figure 3.* Effect of confidence threshold $\tau$ on number of cultural regions in equilibrium for three different noise regimes, $F=5$, $F_n=4$, $Q=15$. Averages based on 20 replications per condition after convergence.

Figure 3 shows that without noise, cultural diversity can be sustained with confidence intervals as wide as 80%, a level that corresponds to the maximal cultural distance that Axelrod assumed for the baseline condition. For confidence levels below $\tau=0.70$, the number of cultural regions in equilibrium approaches the theoretical maximum of 100, indicating a population in which almost all agents are cultural isolates. Inspection of the convergence times until equilibrium shows why. From a random start, confidence intervals below 70% are sufficient to effectively preclude interaction. It becomes extremely unlikely that two neighboring agents will have sufficient cultural overlap to obtain a positive probability of interaction[4]. Figure 3 also shows how this pattern changes as confidence intervals increase. At an 80% confidence interval, corresponding to Axelrod's assumption of overlap in at least one out of five nominal features in the baseline model, the number of cultural regions falls to 35.15, but this is still well above the diversity Axelrod found (20) in a model with only nominal features. Only when confidence intervals reach $\tau=0.9$ or higher does diversity collapse into a single monoculture.

While an 80% confidence interval suggests that bounded confidence has a highly robust effect in sustaining diversity, we should keep in mind that this assumes a world without any noise. Figure 3 also shows that noise greatly reduces diversity under bounded confidence. With mutation, some diversity can be sustained up to a confidence interval as broad as $\tau=0.65$, but for larger intervals, the outcome approaches monoculture. For random interaction, the effect is even more extreme. For $\tau>0.25$, monoculture is the unique equilibrium outcome. Convergence times show that the moderate diversity we obtained with random interaction below $\tau>0.25$, is not an equilibrium state. In this region of the parameter space, the distribution of traits failed to converge within $10^7$ iterations. Had we run the experiment longer, monoculture would have been the ineluctable outcome.

Once we take noise into account, bounded confidence seems much less promising as a robust solution to the puzzle of cultural diversity. Moreover, these limited effects were obtained using a test that may have been too easy, in that we allowed only one metric feature out of five. Holding constant the number of cultural features and the size of the confidence interval, the more features that are metric, the less likely it is that two neighbors with random traits will disagree sufficiently to have no further interaction. Accordingly, we expect that the critical confidence threshold (that is, the maximum interval above which diversity collapses) will drop when the number of nominal features, $F_n$, is reduced, ceteris paribus. To test this, we replicated the no-noise condition of Figure 3 and varied the number of nominal features from 0 to 4, keeping the dimensionality of the state space constant at $F$=5. Figure 4 shows the results.

---

4 Traits are initially uniformly distributed integer values within $[0..Q-1]$. For illustration: with $\tau=0.5$, two independent agents can have an overlap $o \geq 1-\tau$, if they either agree on exactly two nominal features and have a distance of 0.5 or less on the metric feature, or they agree on three or more of the nominal features. The probability for overlap on $k$ nominal features can be obtained via the binominal distribution, and the probability for a distance less than or equal to some $x$ on a metric feature is $1-(1-x)^2$. Overall, this yields a probability of about 0.019 for $o \geq 1-\tau$ with $F=5$, $F_n=4$, $Q=15$. For comparison, in the baseline condition the probability for overlap in at least one feature is about 0.29.

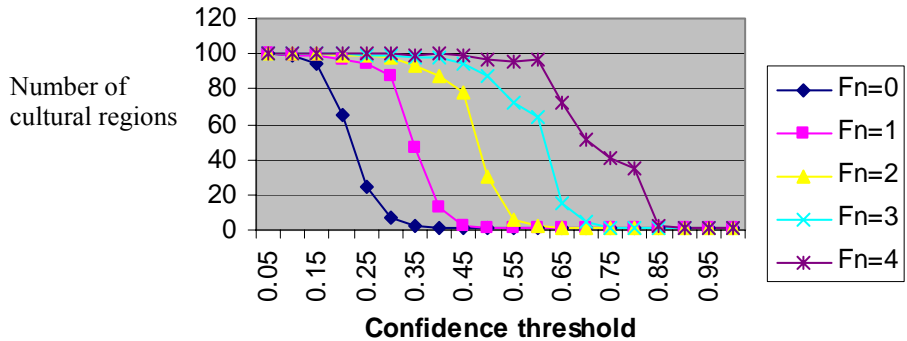

*Figure 4*. Combined effect of confidence threshold $\tau$ and number of nominal features, $F_n$, on number of cultural regions in equilibrium, $F$=5, $Q$=15, $r = r' = 0$. Averages based on 20 replications per condition after convergence.

Figure 4 confirms our expectation. The critical threshold required to sustain stable diversity declines in the number of metric features in the state space, from $\tau$=0.8 with only one metric feature, to $\tau$=0.3 with five metric features. Had we also allowed noise, the results would be even more discouraging.

### 4. Discussion and conclusions

In response to Axelrod (1997), computational models of cultural dissemination have pointed to noise as a key factor that can destabilize cultural diversity and promote cultural consensus. Our study reveals a new source of cultural consensus, metric features. We integrated noise and metric features into Axelrod's original model. Our analyses of the extended cultural dissemination model showed that metric features undermine cultural diversity even without noise. In a metric state space, it is much less likely that neighbors will have zero cultural overlap, leaving a positive probability for ongoing mutual influence, leading ineluctably to monoculture in the long run. Recent extensions of another class of social influence models, opinion formation models, have identified an alternative solution to the puzzle of cultural diversity: bounded confidence. To test this solution, we integrated bounded confidence into our extended model of cultural dissemination. We showed how bounded confidence can sustain stable diversity even in the presence of metric features.

However, our results also raise doubts whether bounded confidence is a sufficient explanation for stable cultural diversity in a world where many cultural features are undoubtedly metric and noise is ever present. If we allow for mutation, random interaction, and the prevalence of metric features, the findings suggest that cultural diversity can only be sustained if confidence intervals are implausibly narrow, such that only agents who are nearly identical are able to influence one another.

Moreover, throughout our study we have assumed very small neighborhoods with minimal overlap. Yet previous work has shown that larger spatial overlap between neighborhoods also increases influence and thus reduces diversity (cf. Axelrod, 1997). Our results may actually be overstating the plausibility of bounded confidence as an explanation for cultural diversity. Given the obvious persistence of cultural diversity, even in an increasingly "small world," the puzzle that motivated our research remains as perplexing as ever.

Looking forward, there are several possible extensions of the cultural dissemination framework that remain neglected by previous studies as well as our own. One or more of these extensions may help to explain persistent diversity. In particular, with the exception of Mark (2003) and Macy et al (2003), previous studies of cultural dissemination have neglected the negative side of homophily and social influence -- xenophobia and differentiation. "Negativity" may explain why agents may remain different even when social interaction between them remains possible. However, Mark does not integrate both positive and negative mechanisms in a single model, and Macy et al. do not incorporate metric features with more than two cultural traits. While we have shown in this paper that metric states greatly reduce cultural diversity under homophily and social influence, we believe it is a promising avenue for future work to integrate xenophobia and differentiation into the cultural dissemination framework that we have elaborated here.